\documentclass[twoside,twocolumn,9pt]{article}
\usepackage{extsizes}
\usepackage[super,sort&compress,comma]{natbib}
\usepackage[version=3]{mhchem}
\usepackage[left=1.5cm, right=1.5cm, top=1.785cm, bottom=2.0cm]{geometry}
\usepackage{balance}
\usepackage{times,mathptmx}
\usepackage{sectsty}
\usepackage{amsfonts}
\usepackage{amsmath}
\usepackage{graphicx}
\usepackage{lastpage}
\usepackage[format=plain,justification=justified,singlelinecheck=false,font={stretch=1.125,small,sf},labelfont=bf,labelsep=space]{caption}
\usepackage{float}
\usepackage{fancyhdr}
\usepackage{fnpos}
\usepackage[english]{babel}
\addto{\captionsenglish}{%
  
}
\usepackage{array}
\usepackage{droidsans}
\usepackage{charter}
\usepackage[T1]{fontenc}
\usepackage[usenames,dvipsnames]{xcolor}
\usepackage{setspace}
\usepackage[compact]{titlesec}
\usepackage{hyperref}
\usepackage{lipsum}
\usepackage{epstopdf}

\definecolor{cream}{RGB}{222,217,201}

\begin{document}

\pagestyle{fancy}
\thispagestyle{plain}
\fancypagestyle{plain}{

\renewcommand{\headrulewidth}{0pt}
}

\makeFNbottom
\makeatletter
\renewcommand\LARGE{\@setfontsize\LARGE{15pt}{17}}
\renewcommand\Large{\@setfontsize\Large{12pt}{14}}
\renewcommand\large{\@setfontsize\large{10pt}{12}}
\renewcommand\footnotesize{\@setfontsize\footnotesize{7pt}{10}}
\makeatother

\renewcommand{\thefootnote}{\fnsymbol{footnote}}
\renewcommand\footnoterule{\vspace*{1pt}%
\color{cream}\hrule width 3.5in height 0.4pt \color{black}\vspace*{5pt}}
\setcounter{secnumdepth}{5}

\makeatletter
\renewcommand\@biblabel[1]{#1}
\renewcommand\@makefntext[1]%
{\noindent\makebox[0pt][r]{\@thefnmark\,}#1}
\makeatother
\renewcommand{\figurename}{\small{Fig.}~}
\sectionfont{\sffamily\Large}
\subsectionfont{\normalsize}
\subsubsectionfont{\bf}
\setstretch{1.125} 
\setlength{\skip\footins}{0.8cm}
\setlength{\footnotesep}{0.25cm}
\setlength{\jot}{10pt}
\titlespacing*{\section}{0pt}{4pt}{4pt}
\titlespacing*{\subsection}{0pt}{15pt}{1pt}



\onecolumn
\noindent\LARGE{\textbf{Vacancies in graphene: an application of adiabatic quantum optimization}} \\

\noindent\large{Virginia Carnevali,\textit{$^{a}$} Ilaria Siloi,\textit{$^{b}$} Rosa Di Felice,\textit{$^{bc}$} and Marco Fornari$^{\ast}$\textit{$^{d}$}} \\

 \noindent\normalsize{Quantum annealers have grown in complexity to the point that quantum computations involving few thousands of qubits are now possible. In this paper, \textcolor{black}{with the intentions to show the feasibility of quantum annealing to tackle problems of physical relevance, we used a simple model, compatible with the capability of current quantum annealers, to study} the relative stability of graphene vacancy defects. By mapping the crucial interactions that dominate carbon-vacancy interchange onto a quadratic unconstrained binary optimization problem, our approach exploits \textcolor{black}{the ground state as well the excited states found by} the quantum annealer to extract all the possible arrangements of multiple defects on the graphene sheet together with their relative formation energies. This approach reproduces known results and provides a stepping stone towards applications of quantum annealing to problems of physical-chemical interest.
} \\




\renewcommand*\rmdefault{bch}\normalfont\upshape
\rmfamily
\section*{}
\vspace{-1cm}


\footnotetext{\textit{$^{a}$~Department of Physics, Central Michigan University, Mt. Pleasant, MI 48859, United States. }}
\footnotetext{\textit{$^{b}$~Department of Physics and Astronomy, University of Southern California, Los Angeles, CA 90089, United States. }}
\footnotetext{\textit{$^{c}$~Center for Quantum Information Science \& Technology, University of Southern California, Los Angeles, California 90089, United States. }}
\footnotetext{\textit{$^{d}$~Department of Physics and Science of Advanced Materials Program, Central Michigan University, Mt. Pleasant, MI 48859, United States. Tel: +1 989 774 2564; E-mail: forna1m@cmich.edu} }



\section{Introduction}
Whether you want to identify the most stable arrangement of water molecules in ice nucleation, the energy barrier of a catalytic reaction, or the stability of a chemical compound, the problem always comes down to finding extrema (minima or maxima) of a specific objective function.
There is no an ideal algorithm that is able to optimize any given objective function. Instead, different methods have been developed, each being suited to a specific class of problems.\cite{Ames} If the first derivative of the objective function is computable, conjugated gradient methods\cite{Desideri} or variable metric methods\cite{Powell} may be used. Alternatively, tabu algorithms,\cite{Ebadi} Metropolis methods,\cite{Haario} or simulated annealing\cite{SA} can be used.
Simulated annealing is a powerful tool to efficiently finding the global minimum of multidimensional objective functions with a large number of local minima.\cite{Finnila}
In simulated annealing the objective function to be minimized is identified with the energy of a statistical-mechanical system. The system is given a temperature as a fictitious control parameter that, through a slow high value,
drives the system to the state with the lowest energy (the ground state of the system).

Quantum annealing (QA) performs a similar task to find the minima (maxima) but it exploits the principles of quantum mechanics.\cite{Finnila,Kadowaki} Indeed, QA is searching simultaneously many configuration space regions thanks to the quantum phenomenon of superposition.\cite{Chancellor} At the beginning of the search, all configurations are equally probable. However, the probability of visiting relevant minima increases  during the annealing process.\cite{Santoro}
Quantum tunneling allows the searching to pass through energy barriers rather than be forced to climb them, reducing the probability of becoming trapped in secondary minima.\cite{Denchev} The role of quantum entanglement in discovering correlations between the configuration space coordinates that lead to global minima has also been discussed.\cite{Lanting}

\textcolor{black}{Due to hardware limitations, current quantum annealers are not suitable yet to treat the full complexity of a physicochemical problem but can solve models appropriately reduced to include fewer relevant degrees of freedom. This work expands recent application of quantum annealing to material science problems:\cite{Molecules,Mulligan,Spectrum,Isomer} traditionally the focus has been on the ground state, in this work we take advantage of the statistical properties of quantum annealing by using a portion of the excited state spectrum of the Ising Hamiltonian. Using a D-Wave quantum annealer we solve a prototypical stability problem in materials science, with the definition of a single objective function that establishes a hierarchy in the configurations. By construction the ground state is identified as a reference point for the energy comparisons and assigns meaning to the excited states.}

\textcolor{black}{The particular stability problem studied in this work regards vacancies in graphene. This has been done using} QA as implemented in D-Wave Systems to explore a potential energy surface\cite{Bunyk,Boixo,Albash} and generate meaningful structural models associated with vacancies in graphene.

D-Wave is an Ising spin matrix with tunable parameters, working at low temperature. The Ising matrix assumes a graph  $C(\mathcal{V},\mathcal{E})$ composed of a set of vertices, $\mathcal{V}$, and edges, $\mathcal{E}$. Each of the $N$ spins is a binary variable located at a vertex. The spins are superconducting flux qubits, and their connectivity is set by the Chimera graph $C(\mathcal{V},\mathcal{E})$.\cite{Serra}
The modern DW\_2000Q\_6 (DW2Q6) has $2041$ physical qubits and operates at $T=13.5 \pm 1 \, mK $.
The limited number of qubits and the limited connectivity imposed by the Chimera graph restrict the application of QA in chemistry and materials science; at the present stage only simple models of larger physical systems can be implemented and solved on DW2Q6.
The model proposed in this work has been constructed to study the stability of graphene vacancies in a free-standing graphene sheet.

Graphene is a purely covalent material where each carbon atom makes three bonds. The removal of one atom breaks three covalent bonds, creating three dangling bonds. A single vacancy corresponds to three non bonded electrons, namely one electron per dangling bond.\cite{Skowron} \citet{Zhang} showed that the formation energies of graphene vacancy defects can be modeled using the information of bond-length and bond-angle distortions, as well as the number of dangling bonds. Dangling bonds are sources of important instabilities for graphene. In response, the host material always tends to minimize the number of dangling bonds by structural reorganization, such as reconstruction and clustering of vacancies.\cite{Lee,Gass,Akhukov,Koskinen,Jafri}
Our simple model looks at the graphene vacancy defects as dangling bonds, which are appropriately arranged by minimizing the energy cost. The complexity of the problem has been limited to match DW2Q6 potentials and does not consider any defects reconstruction contribution that, anyway, may be easily included if higher qubits connectivity becomes available.
The model reproduces the correct order for the stability of graphene vacancy defects as normalized to the number of vacancies, thus promoting a cluster of vacancies rather than isolated vacancies. The model discriminates different arrangements and provides competing and coexisting local structures.
\textcolor{black}{The results are in agreement with more complex theoretical analyses \cite{Banhart,Sorkin,Wang} as well as with experimental measurements \cite{Wang_exp,Carnevali} where, even if graphene is grown over a substrate, the hierarchy in the stability of defect configurations is comparable to the one of free-standing graphene.}

\section{QUBO formulation}\label{secQUBO}
DW2Q6 minimizes objective functions expressed as quadratic unconstrained binary optimization (QUBO) problems (or equivalently as Ising Hamiltonians),\cite{Glover} $Q(\mathbf{x})=\mathbf{x}^T\hat{Q}\mathbf{x}$ where $\mathbf{x}$ is an array of binary variables, and $\hat{Q}$ a positive defined matrix representing the function to be minimized.
In order to implement the QUBO on a physical processor, one has to identify a set of interacting qubits, which represent the off-diagonal terms in the QUBO.\cite{Humble} To compensate for the limited connectivity as represented in the Chimera graph, logical variables are encoded in chains of physical qubits. A chain of physical qubits simulates a binary variable and reproduces the interactions in $Q(\mathbf{x})$. The procedure to define connected logical qubits is known as a minor embedding \cite{Okada}, and the tuning of such a procedure strongly affects the performance of the quantum annealing process.

The protocol proposed in this work includes: 1) mapping the crystal structure into a graph $G(\mathcal{V},\mathcal{E})$, where nodes are the atomic sites and edges are the atomic bonds; 2) defining the binary variables that encode the problem; and 3) designing a QUBO that returns structural model(s) of the desired configuration as ground-state or excited states.

The hexagonal carbon lattice of graphene is generated by two sublattices; considering the coordination of each atom, a graphene sheet including $k$ carbon atoms can be mapped in an order-3 bipartite graph $G(\mathcal{V},\mathcal{E})$ with $|\mathcal{V}|=k$ nodes and $|\mathcal{E}|=3k/2$ edges, where the independent subsets of nodes represent the two sublattices (see Fig. \ref{Vac}-a). Independently of the number of atoms treated in the simulation cell, periodic boundary conditions are considered to simulate the infinite sheet. It is always possible to find a representation of $G(\mathcal{V},\mathcal{E})$ in terms of square adjacency matrix $A_{ij}$ with dimension $k$, where $A_{ij}=1$ whenever node $i$ is connected to node $j$, otherwise $A_{ij}=0$. For the graph in Fig. \ref{Vac}-a (top left panel) the adjacency matrix is
\begin{equation*}
A_{ij} =
\begin{pmatrix}
0 & 0 & 0 & 1 & 1 & 0 & 0 & 0 & 0 & 1 & 0 & 0 \\
0 & 0 & 0 & 0 & 1 & 1 & 0 & 0 & 0 & 0 & 1 & 0 \\
0 & 0 & 0 & 1 & 0 & 1 & 0 & 0 & 0 & 0 & 0 & 1 \\
1 & 0 & 1 & 0 & 0 & 0 & 1 & 0 & 0 & 0 & 0 & 0 \\
0 & 1 & 1 & 0 & 0 & 0 & 0 & 0 & 1 & 0 & 0 & 0 \\
0 & 0 & 0 & 1 & 0 & 0 & 0 & 0 & 0 & 1 & 0 & 1 \\
0 & 0 & 0 & 0 & 1 & 0 & 0 & 0 & 0 & 1 & 1 & 0 \\
0 & 0 & 0 & 0 & 0 & 1 & 0 & 0 & 0 & 0 & 1 & 1 \\
1 & 0 & 0 & 0 & 0 & 0 & 1 & 1 & 0 & 0 & 0 & 0 \\
0 & 1 & 0 & 0 & 0 & 0 & 0 & 1 & 1 & 0 & 0 & 0 \\
0 & 0 & 1 & 0 & 0 & 0 & 1 & 0 & 1 & 0 & 0 & 0 \\
\end{pmatrix}.
\end{equation*}

Vacancies are treated as fictitious atoms: a node $i$ can be occupied by either a carbon ($\alpha_c$) or a vacancy ($\alpha_v$) atom. Each node $i$ is associated with a couple of binary variables $\{x_{i\alpha_c}, x_{i\alpha_v}\}$, such that $x_{i\alpha_c}=1$ when site $i$ is occupied with a carbon, 0 otherwise. The same holds for $x_{i\alpha_v}$. Thus, the representation in terms of binary variables of a graph with two sites, where the first one is a carbon and the second one a vacancy, is $\{1,0,0,1\}$. Overall, $\mathbf{x}$ is a string containing $2|\mathcal{V}|$ binaries that describe the arrangement of atoms (and fictitious atoms) in the simulation cell. To the variables $\alpha_c$ and $\alpha_v$ are associated occupation numbers, $N_{\alpha_c}$ and $N_{\alpha_v}$, such that $N_{\alpha_c}+N_{\alpha_v}=|\mathcal{V}|$. Dangling bonds are represented by an edge linking a node occupied by $\alpha_c$ and to one occupied by $\alpha_v$.
Our QUBO formulation $\hat{Q}$ is designed to reproduce the correct arrangement of (fictitious) atoms in the graphene sheet, such that the ground state $\mathbf{x}_{gs}=\min_{\mathbf{x}}Q(\mathbf{x})$ is the defect-free graphene. Vacancies result in dangling bonds whose number should be minimized \textcolor{black}{and are encoded in the excited states of the QUBO.}
We model this effect by introducing  a  repulsive term $\sum_{i,j=0}^{d-1}A_{ij}x_{i\alpha_{c}}x_{j\alpha_{v}}$, that penalizes configurations containing dangling bonds.

The QUBO of our model reads:
\begin{equation}
  \label{Qvacancy}
  \begin{split}
  Q(\mathbf{x})&=p\sum_{i=0}^{d-1}\left(1-x_{i\alpha_{v}}-x_{i\alpha_{c}}\right)^2 + \sum_{n=0}^{k-1}\left(N_{\alpha_v}+N_{\alpha_c}-x_{i\alpha_{c}}-x_{i\alpha_{v}}\right)^2  +
   \\ &+ \sum_{i,j=0}^{d-1}A_{ij}x_{i\alpha_{c}}x_{j\alpha_{v}},
  \end{split}
\end{equation}
where $p$ is a constant. The first term assures that each node is occupied only by one type of atom; the second term penalizes configurations with occupations different from $N_{\alpha_c}$ and $N_{\alpha_v}$, and the last term is the repulsive one.

\section{\textcolor{black}{QA technical details}}
Quantum annealing (QA) proceeds from an initial Hamiltonian $ H_0 = \sum_i \sigma_i^x $, where a transverse magnetic field is acting on qubit $i$, to a final Hamiltonian $H$, whose ground state encodes the solution of the computational problem under consideration. QUBO problems are implemented on DW2Q6 as Ising spin models arranged on the Chimera graph. To this aim, the graph representing the off-diagonal QUBO terms has to be embedded in the Chimera graph. D-Wave’s API provides a tool called {\tt minorminer} which searches for the optimal (minor)embedding using a heuristic approach. This tool searches for the shortest chains of physical qubits that reproduce the interactions in the QUBO problem. Ideally a chain of qubits behaves as a string of binary variables, thus qubits are ferromagnetically coupled. The choice of intra-chain coupling ($J_F$) is important as it affects the energy spectrum in the dynamics, and the probability of finding the ground state. $J_F$ couplings should be strong enough to avoid chain-breaking without dominating the dynamics.

The QUBO problem is automatically cast into an Ising Hamiltonian:
\begin{equation}
  \label{h_ising}
  H=\sum_{i \in \mathcal{V}} h_i \sigma_i^z + \sum_{(i,j)\in \mathcal{E}} J_{ij}\sigma_i^z \sigma_j^z
\end{equation}
where the local fields $ \lbrace h_i \rbrace $ and couplings $ \lbrace J_{ij} \rbrace$ are programmable within a few percent Gaussian distributed error. The $ \lbrace \sigma_i^z \rbrace $ represents both a binary variable taking values $\pm$ 1, and the Pauli $z$-matrix. Given a spin configuration $ \lbrace \otimes_{i=0}^{N}\sigma_i^z \rbrace $, $H$ is the total energy of the system. Problems submitted to DW2Q6 are automatically scaled so that $h_i \in [-2,2]$ and $J_{ij} \in [-1,1]$.

The evolution is controlled by $ t \in [0,T]$ through two monotonic functions $A(t)$ and $B(t)$ such that $H_{tot} = A(t)H_0 + B(t)H$ with $A(T) = 0$ and $B(0) = 0$.  During an annealing cycle, the magnitude of $H_0$ is gradually reduced to zero, while the magnitude of $H$ is slowly increased from zero. After each annealing cycle D-Wave returns a set of spin values $ \lbrace \sigma_i^z = \pm 1 \rbrace $ that attempts to minimize the energy given by Eq. \ref{h_ising} (a lower energy indicates better optimization). The annealing cycle is repeated to obtain an accurate statistic of the solution, here 100,000 times.  The length of the annealing cycle $T$ is tunable, and its length affects the probability of reaching the correct solution.

\section{Results}
\begin{figure*}
  \includegraphics[scale=0.42]{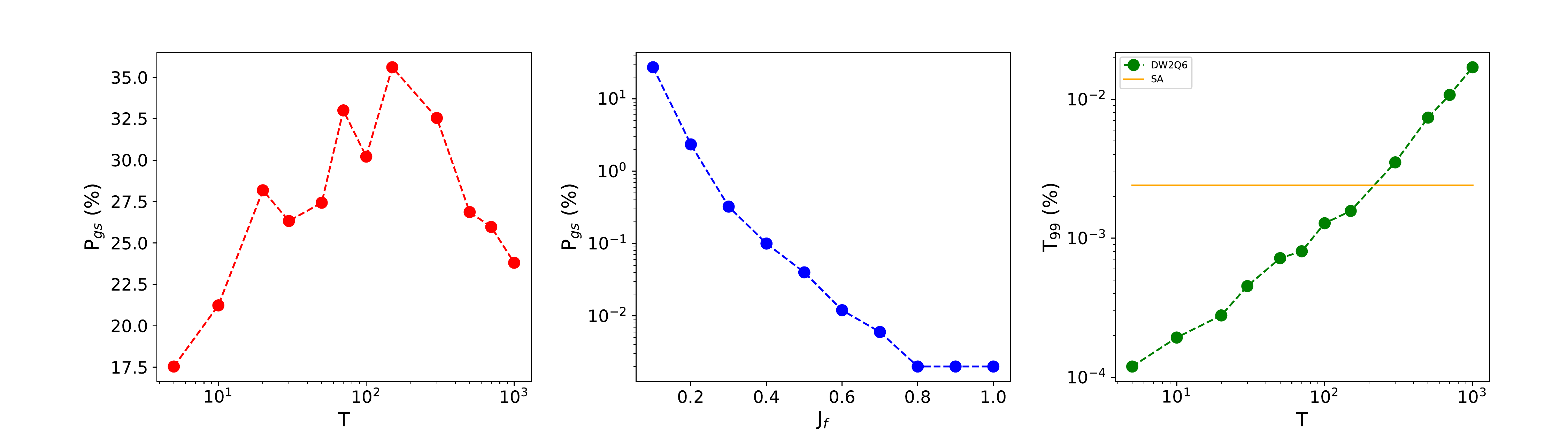}
  \caption{Probability of finding the correct ground state ($P_{gs}$) on DW2Q6 defined as the ratio between the number of correct solutions and total annealing cycles (100,000) as a function of the intra-chain ferromagnetic coupling $J_F$ given in units of the largest coefficient of the embedded Ising model (left panel), and as a function of the annealing time T (middle panel). The right panel reports the comparison for the time performance between DW2Q6 and simulated annealing as a function of the annealing time T. For short annealing times QA is an order of magnitude faster than SA. Only solutions with unbroken chains are counted. Runs are performed using 100 spin reversal operations. For each run on DW2Q6 the embedding was chosen randomly. }
  \label{Pgs}
\end{figure*}

\begin{figure*}
  \includegraphics[scale=1.0]{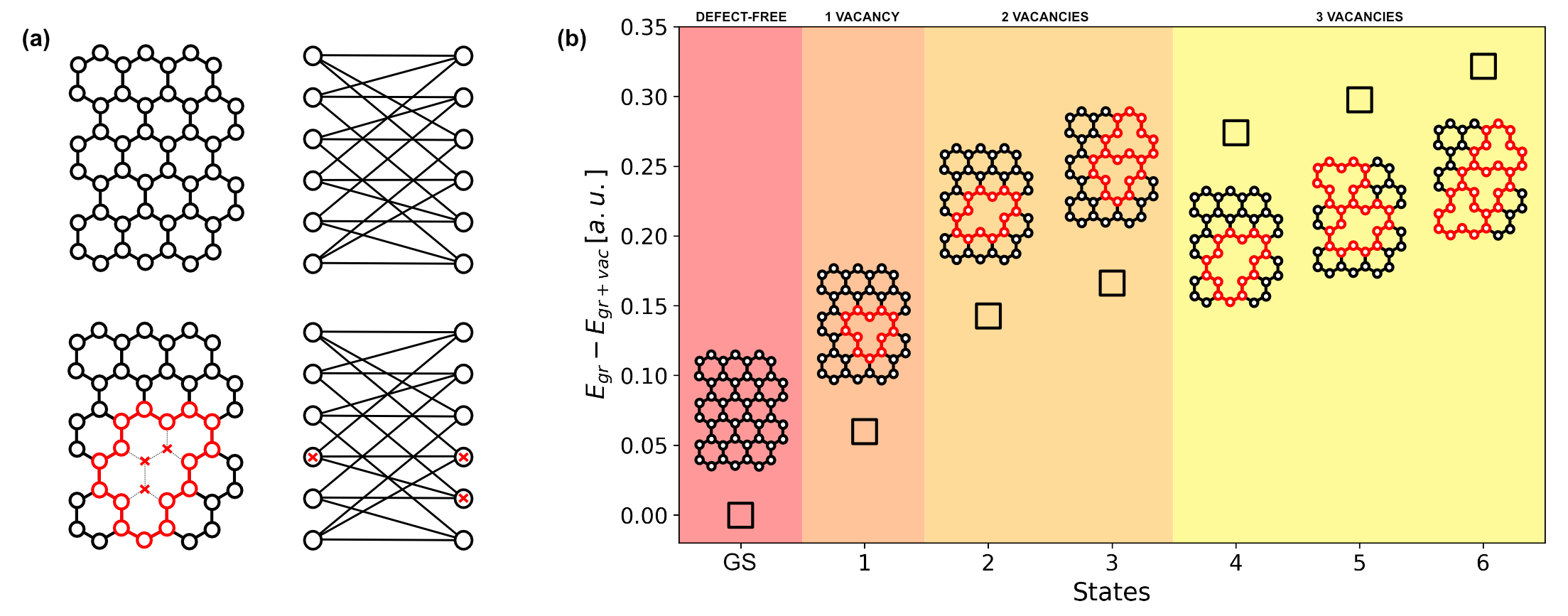}
  \caption{\textbf{(a)} Left panel: stick-and-ball models of free-standing (top) and 3-vacancy defect (bottom) graphene configurations. Carbon atoms are represented in white or in red if bordering a defect, missing carbon atoms are indicated by red crosses, while dangling bonds by black dotted lines. Right panel: representations of defect-free (top) and 3-vacancy defect (bottom) graphene configurations as order-3 bipartite graph with two disjoint parts corresponding to the inequivalent carbon atoms in the Bravais lattice. Carbon atoms at the border of the defect are not highlighted here. \textbf{(b)} Energy spectrum of the QUBO formulation. The ground state coincides with the free-standing graphene configuration; the $1^{st}$ excited state refers to 1-defect graphene configuration; $2^{nd}$ and $3^{rd}$ refer to 2-defect graphene configurations, where the lower in energy ($2^{nd}$) corresponds to the clustered 2-vacancy configuration and the higher ($3^{rd}$) to the configuration made by two separated 1-vacancies; $4^{th}$, $5^{th}$, and $6^{th}$ refer to 3-defect graphene configurations, where, again, the lower in energy correspond to the clustered 3-vacancy configuration. The energy values are reported as differences with respect to the ground state energy. More extended vacancy defects are not considered because of the low probability to be experimentally detected due to their higher instability. For each state, a pictorial representation of the configuration is given. A suitable choice for the QUBO parameters permits DW2Q6 to detect many relevant configurations.}
  \label{Vac}
\end{figure*}

In our QA calculations, it is adopted a simulation cell whose bipartite graph $G(\mathcal{V},\mathcal{E})$ has $|\mathcal{V}|=12$ and $|\mathcal{E}|=36$ (Fig. \ref{Vac}-a, top). This is the smallest cell that allows the accommodation of a 3-vacancy defect structure; extended defects are of less interest for free-standing graphene. The vector $\mathbf{x} \in \mathbb{Z}_2^{2|\mathcal{V}|}$ represents the arrangement of carbon atoms and vacancies in the simulation cell and contains $2|\mathcal{V}|=24$ binary variables. To satisfy the condition of free-standing graphene sheet as ground state, $N_{\alpha_c}=|\mathcal{V}|$ and $N_{\alpha_v}=0$. Thus, the designed $Q(\mathbf{x})$ (see Sec. \ref{secQUBO}) returns by construction the free-standing graphene as the lowest energy configuration.
\textcolor{black}{The reliability on the ground state of the QUBO has been tested with a standard analysis of the probability of finding the ground state (P$_{gs}$). The scope of this analysis is to demonstrate that our QUBO is well designed and to demonstrate that the parameters of DW2Q6 can be tuned to maximize the P$_{gs}$. Besides the tuning of the $J_F$ and the annealing time, it has been shown that the choice of the embedding can affect significantly the P$_{gs}$.\cite{Chinese} Because the main goal of this work is to profit by the statistical properties of quantum annealing and not strictly maximize the P$_{gs}$, the embedding was chosen randomly each run.}

\textcolor{black}{Fig. \ref{Pgs} shows the effect of $J_F$ and annealing time T on P$_{gs}$. The P$_{gs}$ reaches the 27\% with $J_F=0.1$ and decreases with the increasing of the $J_F$. We have also shown that choice of the penalty constant $p$ affects overall the P$_{gs}$, while the best $J_F$ is independent from the choice of $p$ (see Supplementary Material). The analysis on the annealing time T indicates $200\,\mu s$ is the best value for maximizing the P$_{gs}$. The accuracy on the solution depends also on the number of spin reversal (SR) operations which are performed to avoid the effect of possible bias on the physical couplers. We have identified SR=100 as a good value for the convergence of the performance. The average quality of the embeddings and the time performance of quantum annealing are assessed by computing the time to get the solution with probability $99\%$ (T$_{99}$) (see Supplementary Material). Since P$_{gs}$ does not vary significantly with T (see Fig. \ref{Pgs}), the time to solution increases with the annealing time. We compare the performance with simulated annealing as implemented in Ocean \cite{Ocean} (see Supplementary Material) and we observe that DW2Q6 performs an order of magnitude better for the shortest analysis times.}

A graphical representation of a defect-free configuration and 3-vacancy defected graphene sheet are sketched in Fig. \ref{Vac}-a (top and bottom respectively), both as stick-and-ball models (left) and bipartite graphs (right), where carbon atoms are shown as white nodes and fictitious vacancy atoms as red crosses. In the stick-and-ball model the carbon atoms at the border of the defect are highlighted in red and the missing bonds are shown as black dotted lines. The bipartite graphs represent the simulation cell containing 12 atoms with periodic boundary conditions.

By replacing carbon atoms with ``vacancy atoms'', graphene goes from a ground state configuration to an increasingly defected one. Defected configurations correspond to excited states of the QUBO since $Q(\mathbf{x})$ quantifies the deviation from the ground state configuration. These variations are due to the violation of the second term of Eq. \ref{Qvacancy}.
It turns out that the larger the number of vacancies, the higher the corresponding $Q(\mathbf{x})$ values. With the same number of defects, configurations where the vacancies are clustered are preferred with respect to those whom vacancies are spread around the graphene sheet. This adjustment in energy agrees with the minimization of graphene dangling bonds and it is caused by the last term of Eq.\ref{Qvacancy} (Fig. \ref{Vac}-b).

DW2Q6 correctly identifies the ground state of $Q(\mathbf{x})$ and the defected graphene configurations as excited states (Fig. \ref{Vac}-b).
In the implementation, 24 logical variables encoding the arrangement of carbon and fictitious vacancy atoms result in 156 qubits on the Chimera graph, where the longest chains measure 8 qubits. The choice of $p=10$ (Eq. \ref{Qvacancy}) is driven by the need of setting the correct ground state and exploring the portion of excited states relevant to the problem. Larger values of $p$ do not allow the detection of the correct ground state by DW2Q6, while for smaller values some interesting excited states turn out to be too high in energy to be detected during the QA process. \textcolor{black}{We decided to use $J_F=0.6$ and $T=20 \, \mu s$ since the maximization of the P$_{gs}$ is not required for our procedure. Indeed, the goal here is to have access to as many excited state as possible, making irrelevant the parameters that maximize the P$_{gs}$.}
\textcolor{black}{The statistical approach inherent in the QA returns all the defect arrangements in the simulation cell providing information on the degeneracies of structural models. The relative stability hierarchy of the defects shows the defect-free graphene as the most stable configuration, followed by single, double and triple vacancies. A cluster of $N$-vacancies is lower in energy with respect to $N$ separated vacancies (Fig.\ref{Vac}-b).
These solutions of QA reflect the theoretical literature on the topic \cite{Banhart,Sorkin,Wang} and the experimental evidences of graphene vacancy defects in presence of a substrate.\cite{Wang_exp,Carnevali} It is worthy to underline that the hierarchy of the formation energies for single, double and triple vacancies in the free-standing graphene is comparable to the hierarchy of the formation energies of the same defects in graphene grown on a substrate. Because we are dealing with a simplified model where no effects due to the defect reconstruction and electronic contributions are taken into account, the quality of the results is based only on the relative stability of the configurations.}

\section{Conclusions}
We have shown the feasibility of a D-Wave quantum computer in solving physical problems applied to atomistic configuration stability.
\textcolor{black}{In order to accomplish this task, we have developed a new method that exploits the statistical properties of quantum annealing by assigning physical to the full spectrum of the QUBO. In this scenario, the ground state of the QUBO is the natural reference for the energy comparison between different configurations of the system. We believe that this is a stepping stone in material science simulations where setting a reference for the energy comparison is always a non-trivial task.}

The developed method allows the mapping of a stability problem applied to graphene vacancy defect onto a QUBO formulation, making it solvable by DW2Q6. Our model can be generalized to more complex physical problems involving configuration stability. A natural extension of this work could include graphene over a substrate, stability of graphene's registries, \cite{Bianchini} and/or  vacancy defects stability also in presence of adatom coming from the substrate.\cite{Wang_exp,Carnevali} This scenario requires a more complex QUBO formulation in terms of fictitious atoms and a larger simulation cell (larger then a 3-vacancy defect).

In this work, rather than requiring a different simulation's input for each configuration of the system, a single model expressed as a QUBO formulation already provides the necessary information about both pristine and defected systems, streamlining the energy comparison between the configurations of the system. This is a peculiarity of D-Wave that arises from the statistical approach required by QA. In principle, this allows access to all equi-energetic configurations of a multidimensional space with a single QUBO formulation.
Thinking of a future possibility of building larger, more accurate, and fully connected quantum annealers, the approach described in this work opens an interesting and promising prospective in solving more and more complex chemical and physical problems.

\section*{Conflicts of interest}
There are no conflicts to declare.

\section*{Acknowledgements}
This work was supported by the Office of Basic Energy Sciences, US Department of Energy, DE-SC0019432.


\balance

\providecommand*{\mcitethebibliography}{\thebibliography}
\csname @ifundefined\endcsname{endmcitethebibliography}
{\let\endmcitethebibliography\endthebibliography}{}

\end{document}


\pagestyle{fancy}
\thispagestyle{plain}
\fancypagestyle{plain}{

\renewcommand{\headrulewidth}{0pt}
}

\makeFNbottom
\makeatletter
\renewcommand\LARGE{\@setfontsize\LARGE{15pt}{17}}
\renewcommand\Large{\@setfontsize\Large{12pt}{14}}
\renewcommand\large{\@setfontsize\large{10pt}{12}}
\renewcommand\footnotesize{\@setfontsize\footnotesize{7pt}{10}}
\makeatother

\renewcommand{\thefootnote}{\fnsymbol{footnote}}
\renewcommand\footnoterule{\vspace*{1pt}%
\color{cream}\hrule width 3.5in height 0.4pt \color{black}\vspace*{5pt}}
\setcounter{secnumdepth}{5}

\makeatletter
\renewcommand\@biblabel[1]{#1}
\renewcommand\@makefntext[1]%
{\noindent\makebox[0pt][r]{\@thefnmark\,}#1}
\makeatother
\renewcommand{\figurename}{\small{Fig.}~}
\sectionfont{\sffamily\Large}
\subsectionfont{\normalsize}
\subsubsectionfont{\bf}
\setstretch{1.125} 
\setlength{\skip\footins}{0.8cm}
\setlength{\footnotesep}{0.25cm}
\setlength{\jot}{10pt}
\titlespacing*{\section}{0pt}{4pt}{4pt}
\titlespacing*{\subsection}{0pt}{15pt}{1pt}


\onecolumn
\noindent\LARGE{\textbf{Supplementary material: Vacancies in graphene: an application of adiabatic quantum optimization}} \\

\noindent\large{Virginia Carnevali,\textit{$^{a}$} Ilaria Siloi,\textit{$^{b}$} Rosa Di Felice,\textit{$^{bc}$} and Marco Fornari$^{\ast}$\textit{$^{d}$}} \\


\renewcommand*\rmdefault{bch}\normalfont\upshape
\rmfamily
\section*{}
\vspace{-1cm}


\footnotetext{\textit{$^{a}$~Department of Physics, Central Michigan University, Mt. Pleasant, MI 48859, United States. }}
\footnotetext{\textit{$^{b}$~Department of Physics and Astronomy, University of Southern California, Los Angeles, CA 90089, United States. }}
\footnotetext{\textit{$^{c}$~Center for Quantum Information Science \& Technology, University of Southern California, Los Angeles, California 90089, United States. }}
\footnotetext{\textit{$^{d}$~Department of Physics and Science of Advanced Materials Program, Central Michigan University, Mt. Pleasant, MI 48859, United States. Tel: +1 989 774 2564; E-mail: forna1m@cmich.edu} }



\section{DW\_2000Q\_6 hardware}
\begin{figure}[h!]
  \centering
  \includegraphics[scale=0.5]{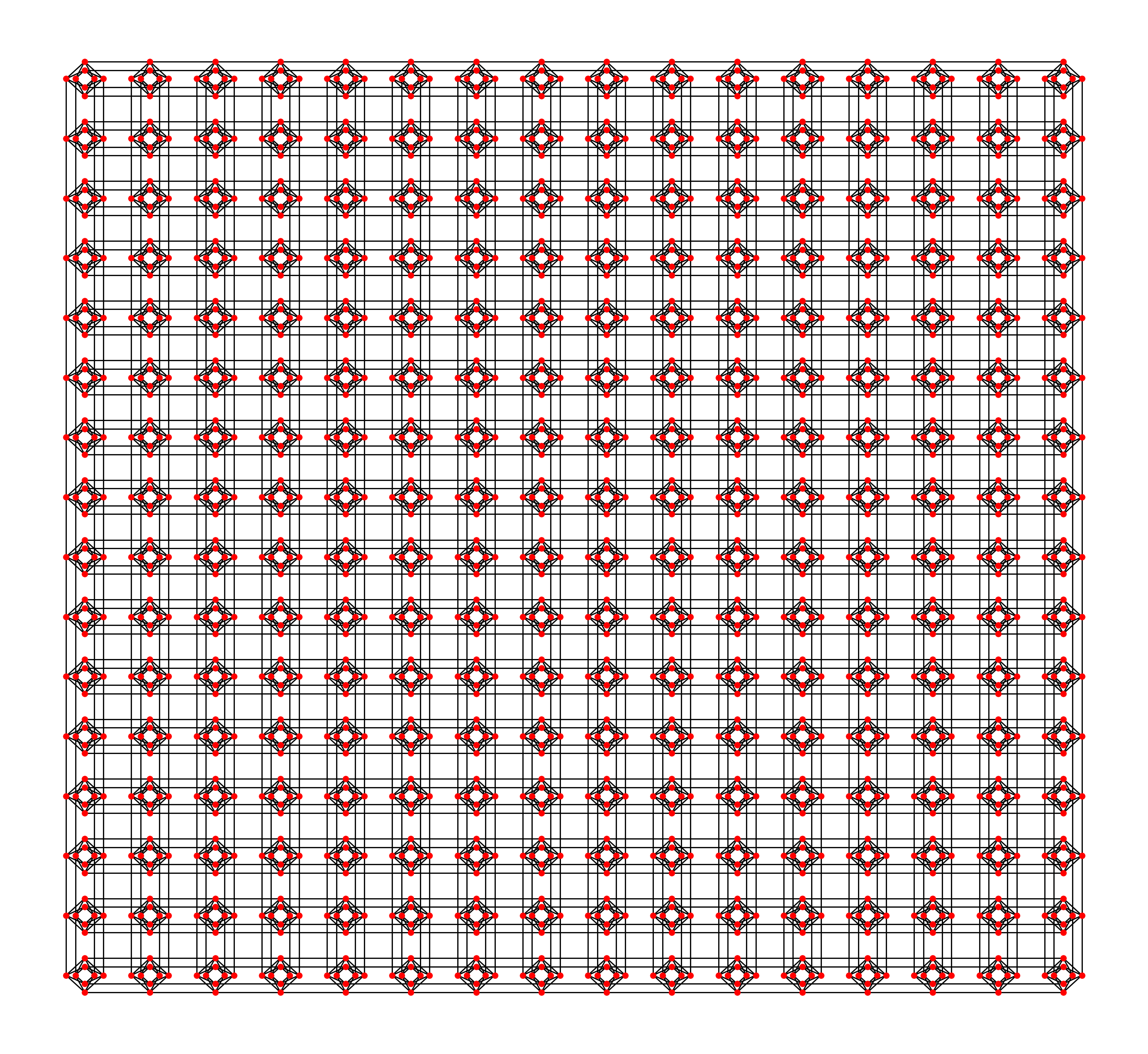}
  \caption{Schematic representation of the fully Chimera graph of DW\_2000Q\_6 (DW2Q6) developed by D-Wave System Inc. and used in this work. Red circles represent qubits while lines represent couplings between qubits. Each qubit can be coupled to a maximum of six other qubits.}
  \label{Chimera}
\end{figure}
In D-Wave quantum annealers, physical qubits are arranged in a Chimera graph $C(\mathcal{V},\mathcal{E})$ topology; the one for DW\_2000Q\_6 (DW2Q6) has N=2041 qubits, and each qubit can coupled with maximum other six qubits (Fig. \ref{Chimera}). Usually, a problem represented by a QUBO is mapped into a graph $G(\mathcal{V},\mathcal{E})$ that has higher connectivity with respect to the $C(\mathcal{V},\mathcal{E})$. This is also the case of our work. The point, then, is mapping (through the so-called minor embedding process) the ``logical problem" on the physical DW2Q6 hardware \cite{Choi,Humble}. This introduce the concepts of ``physical qubits", meaning a set of linked (logical) qubits acting as a single (physical) qubit. D-Wave’s API provides a tool called {\tt minorminer} which heuristically searches for the optimal (minor)embedding \cite{Cai}. {\tt minorminer} searches for the minimum length of each embedding chain, as shorter chains are less prone to breaking into domains of physical qubits with opposite spin orientations. Physical qubits belonging to the same logical qubit are strongly coupled by an intra-chain bias ($J_F$). The $J_F$ is important as it affects the time-dependent energy spectrum in the adiabatic evolution and determines the ability of the chain to act as a single variable; the $J_F$ bias should be strong enough to avoid chain-breaking without dominating the dynamics. Ideally, once the $J_F$ is set correctly, all solutions returned by DW2Q6 would correspond to valid logical solutions, where all the physical qubits within a logical qubit would have the same spin. Because of the probabilistic nature of quantum annealing, as well as noise from different sources, often some percentage of solutions have broken chains. Three approaches are possible to deal with broken chain solutions. The first method is simply discarding these solutions; the second one, so-called ``majority voting", assigns to the logical qubit the spin with the major vote within the physical qubits of the chain; the last one is to go through the broken chains one by one and select the setting that minimizes the energy of the QUBO. The approach used in this work is the first one.

\section{Time To Solution: quantum annealing vs simulated annealing}
The time performance of DW2Q6 is assessed by computing the time to get the solution with probability 99\%:
$$
\mbox{T}_{99}=\dfrac{ln(1-0.99)}{ln(1-\mbox{P}_{gs})} T
$$
where T is the annealing time and P$_{gs}$ is the probability of finding the correct ground state at annealing time T.
Because the best T is different from one embedding to another, it is not possible to select a T that allows top performance for all the embeddings. According to this, we have evaluated $\mbox{T}_{99}$ using the default T=$20\mu s$. The performance of DW2Q6 has been compared also with simulated annealing as implemented in D-Wave Ocean SDK \cite{Ocean}. Following the literature \cite{Tameem_PhysRevX2018,Chinese}, the time to solution for SA reads as:
$$
\mbox{TTS}=N^2\dfrac{ln(1-0.99)}{ln(1-\mbox{P}_{gs})}\tau_{s}n_{s}
$$
where $N$ is the number of logical variables of the problem, $n_{s}$ is the number of sweeps, and $\tau_{s}=1/f_{SA}$ the time required to perform a single sweep ($f_{SA}$ is the number of spin updates per unit time). Because the SA algorithm performs one spin flip per time step and the clock rate of our cpu is $2.4$ GHz, $f_{SA}=2 \,\text{ns}^{-1}$. The chosen SA annealing schedule $\beta$ is comparable to the one of DW2Q6.

\section{Results by Quantum Annealing: penalty constant and intra-chain coupling}
\begin{figure}
  \centering
  \includegraphics[scale=0.6]{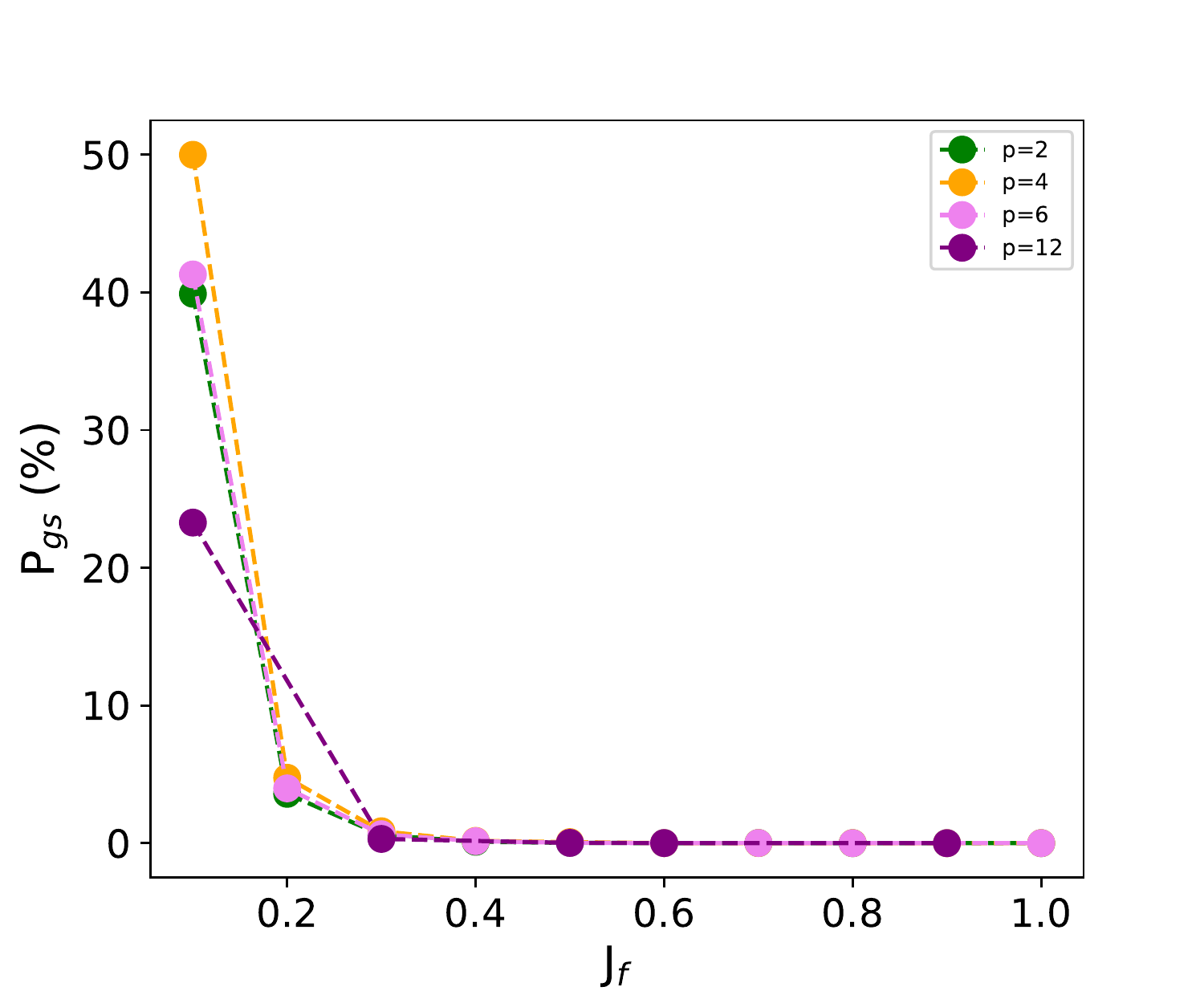}
  \caption{Probability of finding the correct ground state (P$_{gs}$) as a function of the intra-chain coupling ($J_F$) for different penalty constant $p$ (see Eq.1, main text). Only solutions with unbroken chains are counted. Runs are performed using 100 spin reversal operations. For each run on DW2Q6 the embedding was chosen randomly.}
  \label{Pgs_p}
\end{figure}
In this section, we report a comparison in the performance of DW2Q6, measured as probability of finding the correct ground state (P$_{gs}$) as a function of the intra-chain coupling $J_F$. The P$_{gs}$ has been computed for different values of the penalty constant $p$ (see Eq. 1,  main text). The coefficients in front of the penalty functions have a main role in tuning the gap between the ground state and the first excited state of the QUBO \cite{Somma_2013,Choi_2020}. Indeed, we notice that the P$_{gs}$ for quantum annealing decreases with the increasing of $p$ (Fig. \ref{Pgs_p}), while the trend and the best choice for the $J_F$ do not change with $p$.

\providecommand*{\mcitethebibliography}{\thebibliography}
\csname @ifundefined\endcsname{endmcitethebibliography}
{\let\endmcitethebibliography\endthebibliography}{}